\documentclass[]{piparticle-final}
\usepackage{graphicx}
\usepackage{amsmath}

\begin{document}

\volume{1}               % To be inserted by Editor
\articlenumber{010002}   % To be inserted by Editor
\journalyear{2009}       % To be inserted by Editor
\editor{M. C. Barbosa}   % To be inserted by Editor
\reviewers{H. Fort (Universidad de la Rep\'ublica, Uruguay)}  % To be inserted by Editor
\received{6 July 2009}     % To be inserted by Editor
\accepted{2 September 2009}   % To be inserted by Editor
\runningauthor{D. H. Zanette}  % To be inserted by Editor
\doi{010002}         % To be inserted by Editor

\title{A note on the consensus time of mean-field majority-rule dynamics}

\author{Dami\'an H. Zanette\cite{inst1}\thanks{E-mail: zanette@cab.cnea.gov.ar}}

\pipabstract{In this work, it is pointed out that in the mean-field version of
majority-rule opinion dynamics, the dependence of the consensus time
on the population size exhibits two regimes. This is determined by the size
distribution of the groups that, at each evolution step, gather to
reach agreement. When the group size distribution has a finite mean
value, the previously known logarithmic dependence on the population
size holds. On the other hand, when the mean group size diverges,
the consensus time and the population size are related through a
power law. Numerical simulations validate this semi-quantitative
analytical prediction.}

\maketitle

\blfootnote{
\begin{theaffiliation}{99}
\institution{inst1} Consejo Nacional de Investigaciones
Cient\'{\i}ficas y T\'ecnicas, Centro At\'omico Bariloche and
Instituto Balseiro, 8400 San Carlos de Bariloche, R\'{\i}o Negro,
Argentina.
\end{theaffiliation}
}

Much attention has been recently paid, in the context of statistical
physics, to models of social processes where ordered states emerge
spontaneously out of disordered initial conditions (homogeneity from
heterogeneity, dominance from diversity, consensus from
disagreement, etc.) \cite{Loreto}. Not unexpectedly, many of them
are adaptations of well-known models for coarsening in interacting
spin systems, whose dynamical rules are reinterpreted in the
framework of social-like phenomena. The voter model
\cite{Scheucher,Krap} and the majority rule model \cite{Galam,KR}
are paradigmatic examples. In the latter, consensus in a large
population is reached by accumulative agreement events, each of them
involving just a group of agents. The present note is aimed at
briefly revisiting previous results on the time needed to reach
consensus in majority-rule dynamics, stressing the role of the size
distribution of the involved groups. It is found that the growth of
the consensus time with the population size shows distinct behaviors
depending on whether the mean value of the group size distribution
is finite or not.

Consider a population of $N$ agents where, at any given time, each
agent has one of two possible opinions, labeled  $+1$ and $-1$. At
each evolution step, a group of $G$ agents ($G$ odd) is selected
from the population, and all of them adopt the opinion of the
majority. Namely, if $i$ is one of the agents in the selected group,
its opinion $s_i$ changes as

\begin{align}
s_i \to {\rm sign}  \sum_j s_j ,
\end{align}
where the sum runs over the agents in the group. Of course, only the
agents, not the majority, effectively change their opinion. In the
mean-field version of this model, the $G$ agents selected at each
step are drawn at random from the entire population.

It is not difficult to realize that the mean-field majority-rule
(MFMR) dynamics is equivalent to a random walk under the action of a
force field. For a finite-size population, this random walk is
moreover subject to absorbing boundary conditions. Think, for
instance, of the number $N_+$ of agents with opinion $+1$. As time
elapses, $N_+$ changes randomly, with transition probabilities that
depend on $N_+$ itself, until it reaches one of the extreme values,
$N_+=0$ or $N$. At this point, all the agents have the same opinion,
the population has reached full consensus, and the dynamics freezes.

In view of this overall behavior, a relevant quantity to
characterize  MFMR dynamics in finite populations is the consensus
time, i.e. the time needed to reach full consensus from a given
initial condition. In particular, one is interested  in determining
how the consensus time depends on the population size $N$. The exact
solution for three-agent groups ($G=3$) \cite{KR} shows that the
average number of  steps  needed  to reach consensus, $S_c$, depends
on $N$ as

\begin{align} \label{S1}
S_c \propto N \log N ,
\end{align}
for large $N$. The proportionality factor depends in turn on the
initial unbalance between the two opinions all over the population.
The analogy of MFMR dynamics with random walks suggests that this
result should also hold for other values of the group size $G$, as
long as $G$ is smaller than $N$. This can be easily verified by
 solving a rate equation for the evolution of $N_+$
\cite{Loreto}. Numerical results and semi-quantitative arguments
\cite{Tessone} show that Eq. (\ref{S1}) is still valid if, instead
of being constant, the value of $G$ is uniformly distributed over a
finite interval.

What would happen, however, if, at each  step, $G$ is drawn from a
probability distribution $p_G$ that allows for values larger than
the population size? If, at a given step, the chosen group size $G$
is equal to or largen than $N$, full consensus will be instantly
attained and the evolution will cease. In the random-walk analogy,
this step would correspond to a single long jump taking the walker
to one of the boundaries. Is it possible that, for certain forms of
the distribution $p_G$, these single large-$G$ events could dominate the
attainment of consensus? If it is so, how is the $N$-dependence of
the consensus time modified?

To give an answer to these questions, assume that $G$ is drawn from
a distribution which, for large $G$, decays as

\begin{align} \label{pG}
p_G \sim G^{-\gamma},
\end{align}
with $\gamma>1$. Tuning the exponent $\gamma$ of this power-law
distribution,  large values of $G$ may become sufficiently frequent
as to control consensus dynamics.

The probability that at the $S$-th  step the selected group size is
$G\ge N$, while in all preceding steps $G<N$, reads

\begin{align}
P_S = \left( \sum_{G=G_{\rm min}}^{N-1} p_G \right)^{S-1} \sum_{G=
N}^{\infty} p_G,
\end{align}
where $G_{\rm min}$ is the minimal value of $G$ allowed for by the
distribution $p_G$. The average waiting time (in evolution steps)
for an event with $G \ge N$ is thus

\begin{align} \label{S2}
S_w = \sum_{S=1}^\infty S P_S = \left( \sum_{G= N}^{\infty} p_G
\right)^{-1} \propto N^{\gamma -1 },
\end{align}
where the last relation holds for large $N$ when $p_G$ verifies Eq.
(\ref{pG}).

Compare now Eqs. (\ref{S1}) and (\ref{S2}). For $\gamma >2$
(respectively, $\gamma \le 2$) and asymptotically large population
sizes, one has $S_w\gg S_c$ (respectively, $S_w\ll S_c$). This
suggests that above the critical exponent $\gamma_{\rm crit}=2$, the
attainment of consensus will be driven by the asymptotic random-walk
features that lead to Eq. (\ref{S1}). For smaller exponents, on the
other hand, consensus will be reached by the occurrence of a
large-$G$ event, in which all the population is entrained at a
single evolution step. Note that $\gamma_{\rm crit}$ stands at the
boundary between the domain for which the mean group size is finite
($\gamma>\gamma_{\rm crit}$) and the domain where it diverges
($\gamma < \gamma_{\rm crit}$).

In order to validate this analysis, numerical simulations of MFMR
dynamics have been performed for population sizes ranging from
$10^2$ to $10^5$. The probability distribution for the group size
$G$ has been introduced as follows. First, define $G=2g+1$. Choosing
$g=1,2,3,\dots$ ensures that the group size is odd and $G\ge 3$.
Then, take for $g$ the probability distribution

\begin{align}
p_g =\frac{ 1}{\zeta (\gamma)}g^{-\gamma} ,
\end{align}
where $\zeta (z)$ is the Riemann zeta function. With this choice,
$p_G$ satisfies Eq. (\ref{pG}). The average waiting time for a
large-$G$ event, given by Eq. (\ref{S2}), can be exactly given as

\begin{align} \label{S2N}
S_w =\frac{ \zeta (\gamma)}{\zeta (\gamma,1+N/2)} ,
\end{align}
where $\zeta (z,a)$ is the generalized Riemann (or Hurwitz
\cite{Hurwitz}) zeta function. In the numerical simulations, both
opinions were equally represented in the initial condition. The
total number of steps needed to reach full consensus, $S$, was
recorded and averaged over series of $10^2$ to $10^6$ realizations
(depending on the population size $N$).

\begin{figure}[h]
\begin{center}
\includegraphics[width=0.48\textwidth]{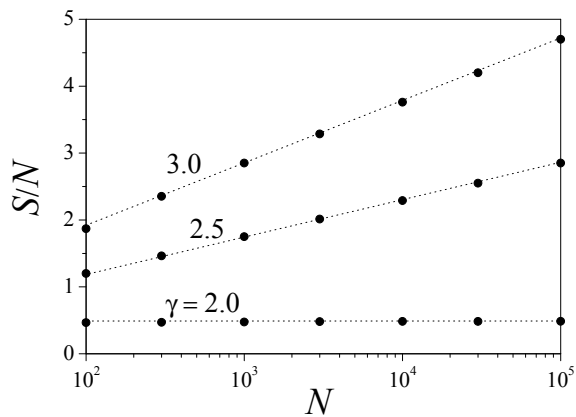}
\end{center}
\caption{Numerical results for the number of  steps needed to reach
consensus, $S$, normalized by the population size $N$, as a function
of $N$, for three values of the exponent $\gamma$. The straight
dotted lines emphasize the validity of Eq. (\ref{S1}) for
$\gamma=2.5$ and $3$. For $\gamma=2$ the line is horizontal,
suggesting $S\propto N$. } \label{fig1}
\end{figure}

The two upper data sets in Fig. \ref{fig1} show the ratio $S/N$ for
two values of the exponent $\gamma > \gamma_{\rm crit}$. Since the
horizontal scale is logarithmic, a linear dependence in this graph
corresponds to the proportionality given by Eq. (\ref{S1}). Dotted
straight lines illustrate this dependence. For these values of
$\gamma$, therefore, the relation between the consensus time and the
population size coincides with that of the case of constant $G$. For
the lowest data set, which corresponds to $\gamma = \gamma_{\rm
crit}$, the relation ceases to hold. The horizontal dotted line
suggests that now $S\propto N$, as predicted for $\gamma=2$ by Eq. (\ref{S2}).

\begin{figure}[h]
\begin{center}
\includegraphics[width=0.48\textwidth]{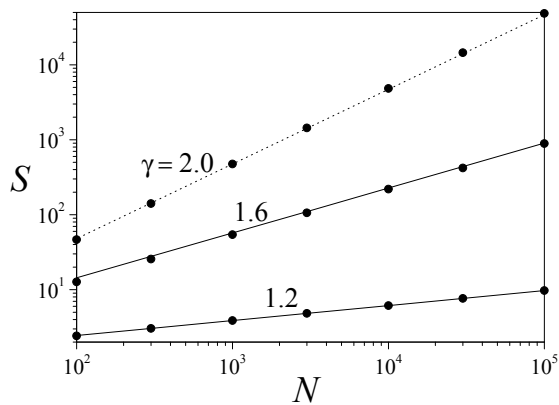}
\end{center}
\caption{Number of steps needed to reach consensus as a function of
the population size, for three values of the exponent $\gamma$. The slope of the straight dotted line equals one. Full curves correspond to
the function $S_w$ given in Eq. (\ref{S2N}).} \label{fig2}
\end{figure}

The log-log plot of Fig. \ref{fig2} shows the number of  steps to
full consensus as a function of the population size for three
exponents $\gamma \le \gamma_{\rm crit}$. The dotted straight line
has unitary slope, representing the proportionality between $S$ and
$N$ for $\gamma=2$. For lower exponents, the full curves are the
graphic representation of $S_w$ as given by Eq. (\ref{S2}). The
excellent agreement between $S_w$ and the numerical results for $S$
demonstrates that, for these values of $\gamma$, the consensus time
in actual realizations of the MFMR process is in fact dominated by
large-$G$ events.

\begin{figure}[h]
\begin{center}
\includegraphics[width=0.48\textwidth]{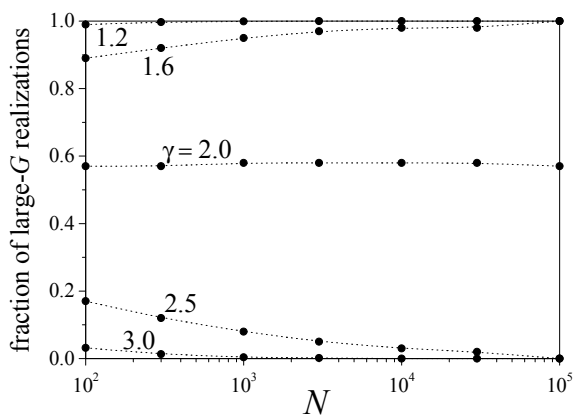}
\end{center}
\caption{Fraction of realizations where consensus is attained
through a large-$G$ event as a function of the population size, for
several values of the exponent $\gamma$.} \label{fig3}
\end{figure}

A further characterization of the two regimes of consensus
attainment is given by the fraction of realizations where consensus
is reached through a large-$G$ event. This is shown in Fig.
\ref{fig3} as a function of the population size. For $\gamma <
\gamma_{\rm crit}$,  consensus is the result of a step involving the
whole population in practically all realizations. As $N$ grows, the
frequency of such realizations increases as well. The opposite
behavior is observed for $\gamma > \gamma_{\rm crit}$. For the
critical exponent, meanwhile, the fraction of large-$G$ realizations
is practically independent of $N$, and fluctuates slightly around
$0.57$.

In summary, it has  been shown here that in majority-rule opinion
dynamics, the dependence of the consensus time on the population size
exhibits two distinct regimes. If the size distribution of the
groups of agents selected at each evolution step decays fast enough, one reobtains the logarithmic analytical result for constant group sizes. If, on the other hand, the
distribution of group sizes decays slowly, as a power law with a
sufficiently small exponent, the dependence of the consensus time on
the population size is also given by a power law. The two regimes
are related to two different mechanisms of consensus attainment: in
the second case, in particular, consensus is reached during events
which involve the whole population at a single evolution step. The
logarithmic regime occurs when the mean group size is finite, while
in the power-law regime the mean value of the distribution of group
sizes diverges. In connection with the random-walk analogy of
majority-rule dynamics, this is reminiscent of the contrasting
features of standard and anomalous diffusion \cite{Levy}.

%\begin{acknowledgements}
%Acknowledge supporting agencies, colleagues, and institutions that provide employment, scholarships or infrastructure.
%\end{acknowledgements}
\raggedbottom
\pagebreak

\end{document}